
\magnification 1200
\tolerance 10000
\baselineskip .5cm
\input epsf.sty
\def\b{\bigskip}
\def\m{\medskip}

\def\iso{$^7$Li/$^6$Li~}
\def\isop{($^7$Li/$^6$Li)}
\def\ab{($^7$Li/H)}

\def\li{$^7$Li}
\def\lis{$^6$Li}
\def \zo{$\zeta$~Oph}
\def \ro{$\rho$~Oph}
\def \zp{$\zeta$~Per}
\def \res{$\lambda/\Delta\lambda$}
\def \ki2{$\chi^2$}
\def\no{\noindent}
\def\ts{~}
\def\ga{\mathrel{\mathchoice {\vcenter{\offinterlineskip\halign{\hfil
$\displaystyle##$\hfil\cr>\cr\sim\cr}}}
{\vcenter{\offinterlineskip\halign{\hfil$\textstyle##$\hfil\cr>\cr\sim\cr}}}
{\vcenter{\offinterlineskip\halign{\hfil$\scriptstyle##$\hfil\cr>\cr\sim\cr}}}
{\vcenter{\offinterlineskip\halign{\hfil$\scriptscriptstyle##$\hfil
\cr>\cr\sim\cr}}}}}
\def\la{\mathrel{\mathchoice {\vcenter{\offinterlineskip\halign{\hfil
$\displaystyle##$\hfil\cr<\cr\sim\cr}}}
{\vcenter{\offinterlineskip\halign{\hfil$\textstyle##$\hfil\cr<\cr\sim\cr}}}
{\vcenter{\offinterlineskip\halign{\hfil$\scriptstyle##$\hfil\cr<\cr\sim\cr}}}
{\vcenter{\offinterlineskip\halign{\hfil$\scriptscriptstyle##$\hfil
\cr<\cr\sim\cr}}}}}

\font\tenrm=cmr10

\font\ninerm=cmr9
\font\ninebf=cmbx9

{\parindent = 0pt
\leftskip = 0pt plus 1fil
\rightskip = 0.5in
\parfillskip = 0pt
IAP--477

\noindent astro-ph/9410055

}
\vskip 2.0cm
\centerline{\bf The Interstellar $^7$Li/$^6$Li Ratio}

\vskip 0.8cm

\tenrm
\centerline{Martin Lemoine$^{1,2}$, Roger Ferlet$^1$}
\vskip 1.2cm

\centerline{1 Institut d'Astrophysique de Paris, CNRS, 98 bis boulevard Arago,
75014 Paris, France.}
\medskip

\centerline{2 D\'epartement d'Astrophysique Relativiste et de Cosmologie,
CNRS,}
\centerline{Observatoire de Meudon, 92195 Meudon C\'edex, France.}
\vskip 2cm
{\ninerm{\ninebf\noindent Abstract.}
We discuss the observational status of the interstellar lithium
isotopic ratio and its significance with respect to the galactic evolution of
lithium.}
\tenrm
{\vfill
\no Based on invited talks given at:

{\leftskip=2cm \no ESO/EIPC Workshop on ``Light Elements Abundances'', Isola
d'Elba, Italy, May 22--28, 1994

\no ESO Workshop on ``Science with the VLT'', Garching, Germany, June 28 --
July
01, 1994

\no IAU General Assembly, Joint Discussion 11, Den Haag, The Netherlands,
August
22, 1994

}}
\eject

\noindent
{\bf 1. Introduction.}
\m
This talk discusses the observational status of the lithium isotopic ratio in
the
interstellar medium (ISM). Although only four such measurements have been
performed up to now, this discussion justifies itself in that the value of
the interstellar \iso ratio may have some drastic consequences on the
galactic evolution of lithium. Moreover it will be shown that the published
values of the interstellar \iso ratio cannot be considered
as representative values for the ISM. In a first part, we therefore
review, discuss, criticize and, where possible, correct each of those
measurements in order to draw a preliminary observational
status of this quantity. We then present the analysis and the results of new
high quality observations of this ratio on a previously observed line of sight,
and discuss the extreme consequences of the atypical values derived. A complete
observational status of the interstellar \iso ratio and its significance with
respect to the galactic evolution of lithium are given as a conclusion.
Before proceeding with this review however, let us briefly outline the purpose
and basic significance of interstellar \iso ratio measurements.

Lithium--7 is now generally accepted to originate in the hot Big Bang
primordial
nucleosynthesis (BBN), with a primordial abundance \ab$\simeq10^{-10}$ (Smith
et al., 1993), in
excellent agreement with the observed uniformity of the \li~abundance in very
metal deficient Pop II stars \ab=1.4$\times10^{-10}$ (Spite et al., 1993).
During the galactic evolution, both lithium isotopes are created by spallation
reactions of galactic cosmic rays (GCR) interacting with the ISM, that yield
\ab$\simeq2.\times10^{-10}$ in 10 Gyrs, with a production ratio
\isop$_{GCR}$=1.4 (Meneguzzi et al., 1975; Reeves, 1994). The major problem in
understanding the
evolution of lithium in the Galaxy is to explain the observed Pop I abundance
of
\li, \ab$_{Pop I}\sim10^{-9}$, of which only 30\% is accounted for by BBN and
GCR
spallation mechanisms, as well as the high \iso ratio measured in
meteorites, representative of the solar system formation epoch 4.6
Gyrs ago, \isop$_{\odot}$=12.3, whereas the above mechanisms predict a
ratio around 2.
In order to account for these two quantities, the existence of an extra source
of \li~of stellar origin has been suggested, AGB C and S stars being the best
candidates as they are observed to be super Li--rich (Abia et al., 1993; Abia
et
al., 1994; see also Mowlavi, 1994, for the production of \li~in AGB stars). It
was argued by Reeves (1993) from the comparison of the meteoritic
\isop$_{\odot}$
ratio and the production rates ratio in GCR spallation reactions that GCR
spallation alone tends to decrease the \iso ratio with time, and that, starting
4.6 Gyrs ago with a ratio \iso=12.3, one should observe today an interstellar
ratio \iso$\simeq$5--6 without production of \li~in stars, or \iso$\ga$6 with
stellar production of \li. Measuring the interstellar \iso ratio thus provides
a
key test for the model of galactic lithium evolution. If this
ratio is found to be \iso$\la$5, then another scenario would have to be
considered; one way out would be to consider a primordial abundance
\ab$\ga10^{-9}$ together with some form of internal mixing that could very well
reproduce the observed ``Spite plateau'' in Pop II stars (Pinseonneault et
al., 1992) and some rotational mechanisms to reproduce {\it via} a gradual
depletion of lithium in stars the observed abundances of \li~in stars of
different
metallicities (Vauclair, 1988). As to now however, there is no obvious way of
reproducing a primordial abundance as high as 10$^{-9}$ since the inhomogeneous
nucleosynthesis models involved up to a few years ago to yield such abundances
no
longer do (Reeves, 1994; Thomas et al., 1994).
\vskip 1.0cm
\no{\bf 2. Previous Measurements}
\m
The fact that there are so few measurements of such an important quantity as
the
interstellar \iso ratio is simply due to the difficulty one has to face to
detect the \lis~isotope. The only accessible resonance lines of lithium form a
transition doublet of \li I at 6707.761--6707.912\ts\AA, with a similar
doublet for \lis I redshifted by 0.160\ts\AA. In regards of the
low abundance of lithium and the nearly complete ionisation of lithium to LiII
in the ISM, the strongest equivalent widths observed for the 6708\ts\AA~line
of \li I are of a few m\AA~on lines of sight already comprising a hydrogen
column density well over 10$^{20}$\ts cm$^{-2}$ (White, 1986). The structure of
the transition is such that the main component of the \lis I doublet is
completely
superimposed on the weaker component of the corresponding \li I doublet.
Therefore, the only absorbing line of \lis~that one may hope to detect is the
weaker one. Assuming a
ratio \iso$\simeq10$, the oscillator strengths ratio of the doublet components
being 2, the typical equivalent width of this line is $\sim30$\ts$\mu$\AA,
which
corresponds to a resolution element of 67\ts m\AA~(i.e. a resolving power
\res=10$^5$) lowered from the continuum by 0.04\%... Moreover, on lines of
sight
showing a column density N(HI)$\ga10^{20}$\ts cm$^{-2}$, several interstellar
components separated by $\sim$5\ts km.s$^{-1}$ are often present so that
the resulting absorption profile of the LiI transition doublet may
be very complex (the isotopic shift of the doublets is 7.2\ts km.s$^{-1}$).

A simulation of this resulting profile corresponding to our best fit solution
for
the \zo~line of sight is shown in Fig.1 as the solid line, the individual
contributions for \li~and \lis~of each of the two interstellar absorbing clouds
being shown in dashed line. Obviously, in
order to detect \lis~and derive a \iso ratio as accurate as possible, one needs
to observe a bright target with a line of sight as dense as possible and whose
velocity structure must be as trivial as possible. It is also necessary to
observe hot stars so that the lithium region be not contaminated by stellar
lines of other elements. Naturally, the number of such targets comes down to
a few. One needs also to reduce the data with care in order to avoid any
instrumental pollution of the profile. Finally, it is necessary to use
sophisticated profile fitting methods in order to probe the observed profile
for
all the contributions shown in Fig.1, since neglecting one of them would mean
neglecting a contribution of the order of the \lis~absorption one is looking
for.

\vbox{
\epsfysize= 8 cm $$\epsfbox[72 130 550 600]{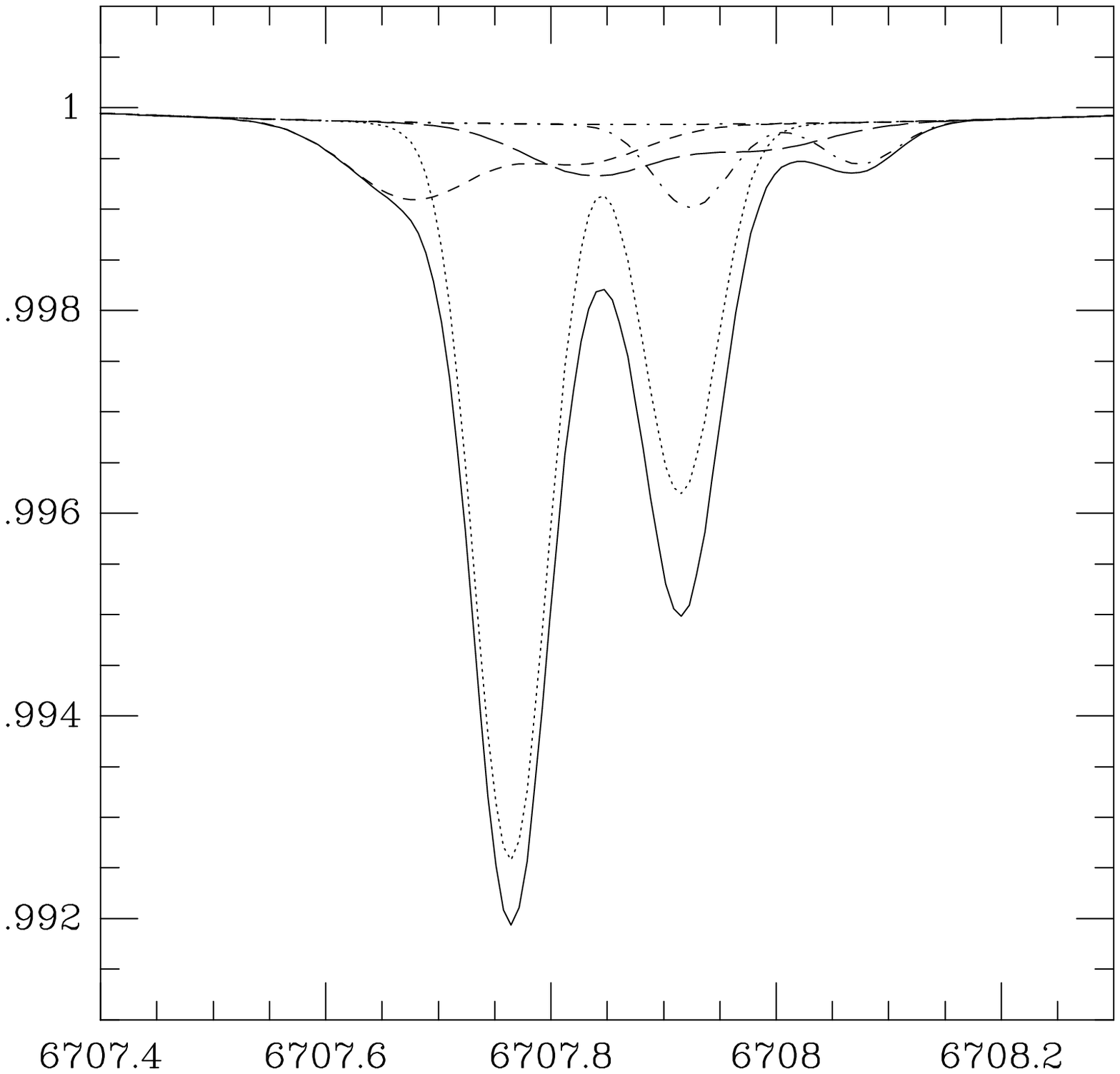}$$

\noindent{\ninerm {\ninebf Fig.1:} Simulation of the observed profile of the
$\lambda6708$\ts\AA~LiI
line as taken from our best fit solution in the direction of \zo~(solid line).
The individual contributions of the two absorbing clouds A and B are shown as:
dotted line \li$_A$, short dash \li$_B$, dot--dash \lis$_A$, long dash
\lis$_B$.
Note that the profile in solid line would only be observed if the
S/N and the sampling were already infinite.}
\bigskip}

Nevertheless, the first estimation of the interstellar \iso ratio was attempted
ten years ago by Ferlet \& Dennefeld (1984, hereafter FD) in the
direction of $\zeta$ Oph using
the Coud\'e Echelle Spectrometer (CES) at the ESO 1.4m Coud\'e Auxiliary
Telescope and a Reticon detector. Although they obtained high quality data
with a resolving power \res=10$^5$ and a signal-to-noise ratio per pixel
S/N=4000, \lis~was
actually not detected for that time. We will therefore not discuss further
the value obtained for the main absorbing cloud:
$$ ^7{\rm Li}/^6{\rm Li}\,\ga\,25\;(\sim38)$$
The first actual detection of \lis~was reported by our group (Lemoine et al.,
1993, hereafter L93) in the direction of \ro. The observations were conducted
at
ESO using the 3.6m Telescope linked via fiber optics to the CES, providing a
resolving power \res=10$^5$ and a signal-to-noise ratio S/N$\sim4000$ per pixel
of a CCD detector.
The KI $\lambda7699$\ts \AA~line was also observed since LiI and
KI are known to behave similarly in the ISM (White, 1986). The typical
equivalent width of the KI line is also much stronger than that of the LiI line
at 6708\ts\AA, and thus allows to derive an accurate velocity structure of the
line of sight. Two interstellar absorbing clouds, one main (A) and one weak (B)
were detected, but only the \iso ratio in cloud A was evaluated, yielding:
$$ ^7{\rm Li}/^6{\rm Li}\,=\,12.5^{+4.3}_{-3.4}\;\;(2\sigma) $$
This value was immediately interpreted by Reeves (1993) as strong evidence for
the existence of an extra source of \li. The only critics we would address to
this measurement is that, due to numerical complexity and to the weakness of
the
B component in \li, the \lis$_B$ contribution was neglected. Using a more
sophisticated profile fitting algorithm (to be shortly discussed in 3.), we
were
able recently to re-analyze this line of sight, and the following ratios were
derived taking into account all detected contributions:
$$ \left(^7{\rm Li}/^6{\rm Li}\right)_A\,=\,11.1$$
$$ \left(^7{\rm Li}/^6{\rm Li}\right)_B\,\sim\,3$$
The error bar on the \iso$_A$ ratio is to remain as previously, i.e. $\pm2$ at
1$\sigma$, but the \isop$_B$ ratio is uncertain since \lis$_B$ is not formally
detected above the photon noise. The fit is shown in Fig.2, the corresponding
\ki2~is 37.4/43 giving a level of confidence of 71\%.

Meyer, Hawkins \& Wright (1993, hereafter MHW) reported shortly thereafter
\iso=6.8$^{+1.4}_{-1.7}$ toward \zo, and \iso=$5.5^{+1.3}_{-1.1}$ toward
$\zeta$ Per, values which cast severe doubt as to the existence of a stellar
source of \li, hence on the ``canonical'' scenario for the galactic evolution
of
lithium. The data were obtained at the KPNO 0.9m telescope at a high quality,
with \res$\simeq$2--3$\times10^5$ and S/N$\simeq2000$ per pixel. These results
seem to us to be questionable for the following reason. It is obvious from the
MHW
spectra that at least two absorbing clouds are well detected in \li~toward each
of the two targets, and only the main component was taken into account in the
profile fitting in each case. It is in fact well known that toward \zo~and
\zp, several interstellar components are indeed present (Welty et al., 1994).
Recalling the discussion above, these values seem to be biased, and are at most
average values of the \iso~ratio on the lines of sight, hence {\it a priori}
not
representative of the general ISM. It happens moreover that the new data we
discuss in 3. were also obtained in the direction of \zo. We derive a ratio
\iso$\simeq$9.8 when
looking for a one cloud solution, with a \ki2~of 303/13 and 103/58 for the KI
line and the LiI line respectively, i.e. respective confidence levels of 0\%
and
0\% because of the non-negligible presence of a second absorbing cloud. This
result is in slight agreement with that derived by MHW albeit higher,
suggesting
that an average value for the \iso ratio on this line sight would lie closer to
8--9 than to 6--7. For similar reasons, we are led to believe that the value
derived toward \zp~is not representative either of the \iso ratios on this line
of sight. These data are of a such quality that they would deserve being
re-analyzed in more detail, as we did for \ro.

\vbox{
\epsfysize= 8 cm $$\epsfbox[72 130 550 600]{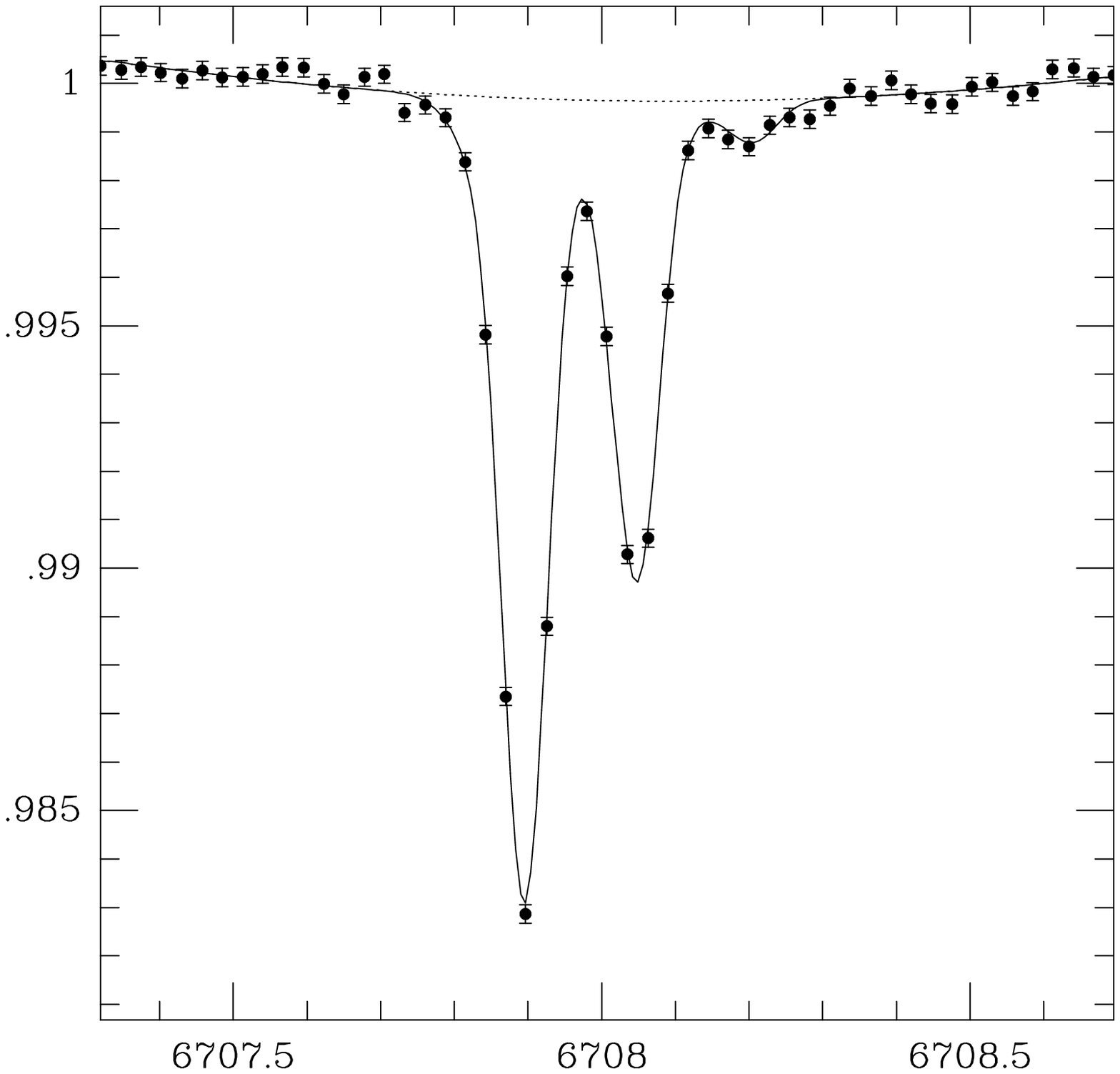}$$

\noindent{\ninerm {\ninebf Fig.2:} Best fit solution for the \ro~line of sight
including all \li~and \lis~contributions for two interstellar absorbing
components. Error bars are 1$\sigma$. The level of confidence of the fit is
71\%.}
\bigskip}

In the end, among these four measurements of the interstellar \iso ratio, it
seems that only the values derived and corrected toward \ro~may be kept, i.e.
\iso$\simeq11.1\pm2$, with an uncertain value \iso$\sim3$ to be confirmed or
not. The values derived by MHW do not seem to be
representative, and the data should be re-analyzed in more detail, as their
high
spectral resolution and high signal-to-noise ratio would for sure allow
accurate
estimations of probably two \iso ratios on each of the two lines of sight.
\vskip 1.0cm
\no{\bf 3. New data toward $\zeta$ Oph}
\m
We obtained new data of the $\lambda$6708\ts\AA~LiI line toward \zo~at the ESO
3.6m
Telescope linked to the CES with fiber optics. \zo~had already been observed by
FD and MHW, yielding two very different values for the \iso ratio. Our data
revealed to be of a higher quality, showing
\res=10$^5$ together with S/N=7500 per pixel. The data were carefully reduced
using different approaches in order to estimate the importance of systematics
on
the LiI profile, hence on the \iso ratios. Indeed, it was found that the noise
is
dominated by interference fringes remnants at a level of $\sim10^{-4}$ {\it
rms}
and not by the photon noise, as it can be seen from the non-gaussian statistics
of the continuum in the spectrum of Fig.3. The
$\lambda7699$\ts\AA~KI line was also observed in order to link our LiI
observations with KI and with others. The spectra were normalized and the
continua fitted using a cross-validation statistical criterion. In order to
probe the profiles for the different contributions (see Fig.1), a
sophisticated profile fitting technique based on a simulated annealing
minimization algorithm allowing to include all kinds of constraints on the
minimization procedures, e.g. atomic and physical constraints on the parameters
that define the profile, was developed. For further details on the reduction
and
analysis of this line of sight, the reader is referred to Lemoine et al.
(1994).
Two interstellar components were detected in KI and in LiI, one main (A) and
one
secondary (B). The limiting
detectable equivalent width of the spectrum is 18\ts$\mu$\AA, or 50\ts$\mu$\AA~
including systematics. We were thus able to derive two \iso ratios:
$$ \left(^7{\rm Li}/^6{\rm Li}\right)_A\,=\,8.6\;\pm0.8\;(\pm1.4)$$
$$ \left(^7{\rm Li}/^6{\rm Li}\right)_B\,=\,1.4\;^{+1.2}_{-0.5}\;(\pm0.6)$$
where the error bars within brackets are associated to systematics and were
estimated by fitting different sets of spectra reduced using different
techniques. In fact, these numbers constitute upper limits to the systematics
for
they somehow include statistical errors associated with the photon noise. The
fit is shown in Fig.3, the corresponding \ki2~is 50.5/54, or equivalently a
level of confidence of 61\%. In Fig.4 (Fig.5), we show the \lis$_A$ (\lis$_B$)
doublet as calculated by subtracting to the observed profile all the calculated
absorption except that of the \lis$_A$ (\lis$_B$) doublet.
As toward \ro, we find an extremely atypical \iso ratio. Why is it that this
ratio be so low?

It might be suggested for instance that the ``strong'' \lis$_B$
contribution observed is due to an instrumental effect. A mimicking absorption
feature, such as an interference fringe remnant as observed in emission on the
blue and red sides of the doublet (see Fig.3), would indeed provide more
\lis$_B$
absorption than real and thus imply a low \iso ratio. However,
as it may be seen from Fig.5, the \lis$_B$ doublet is detected above the noise,
and appears at the right position with seemingly good oscillator strengths
ratio
and fine structure shift of the lines. This makes the presence of an extra
instrumental feature very improbable. Other systematic effects, such as the
contamination of the profile by cosmics, were carefully looked for in
individual
spectra, and none was found.

\vbox{
\epsfysize= 8 cm $$\epsfbox[72 130 550 600]{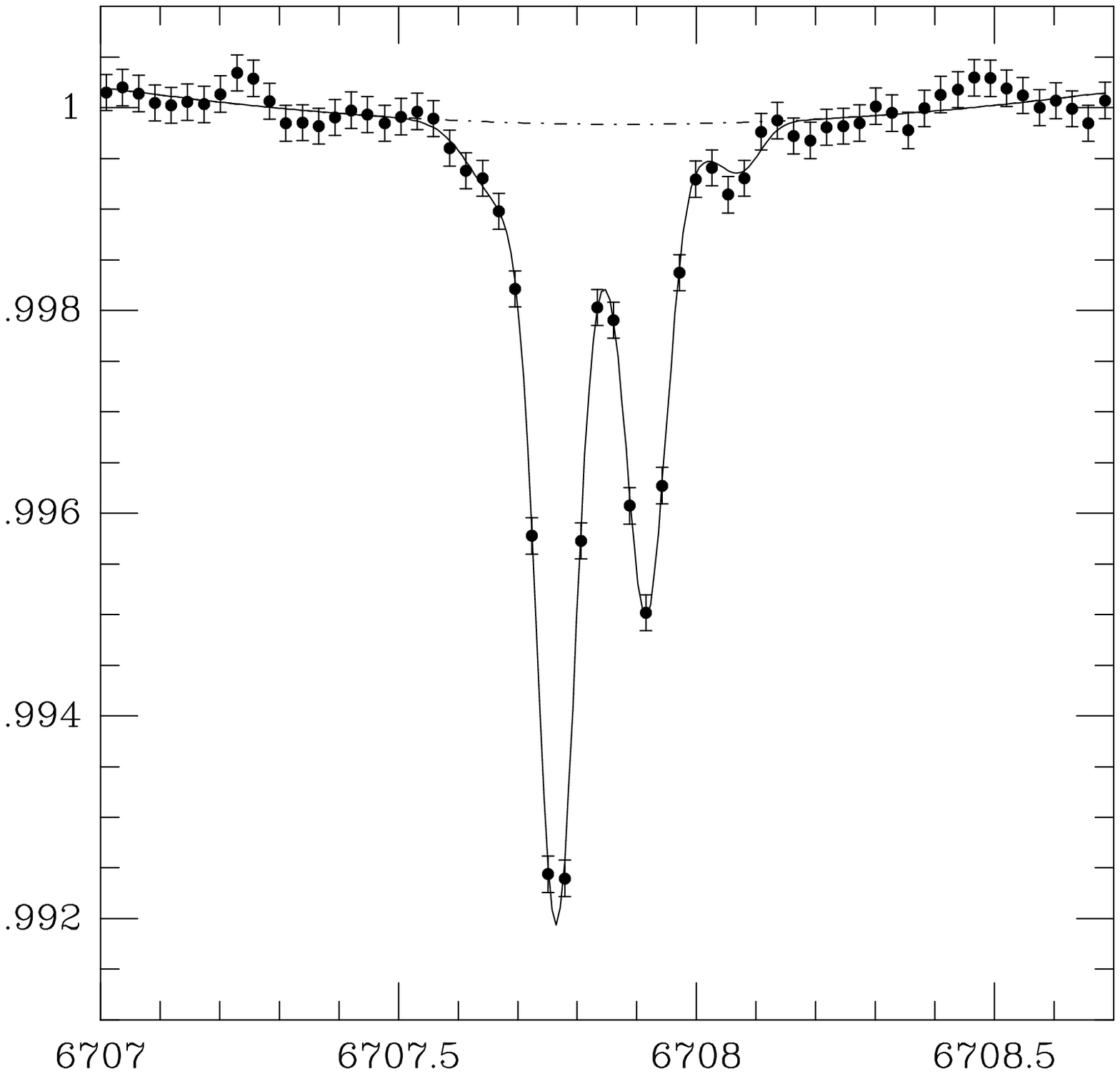}$$

\noindent{\ninerm {\ninebf Fig.3:} Best fit solution for the \zo~line of sight,
including two absorbing components. The level of confidence is now 61\%.}
\bigskip}

Another way to excuse this atypically low ratio would be to assume the presence
of a high radial velocity cloud absorbing in \li I at the position of the
\lis$_B$ ratio, which would mimic perfectly the doublet
structure of the line. This may be
checked by introducing in the observed KI profile an absorption redshifted
so as to correspond to a $^7$LiI$_C$ doublet matching the $^6$LiI$_B$
absorption (see L93). This in fact partly possible from our KI spectrum since
the
redshifted absorption takes place in the red wing of the KI profile (see
Fig.10b, Lemoine et al., 1994). This is not the case for cloud A, where
a similar contamination of the \lis$_A$ doublet is strictly ruled out. It is
well known that many absorbing clouds are present on the \zo~line of sight (see
Welty et al., 1994, NaI observations), but as to whether these other absorbing
clouds contribute in \li I is
another matter. Only extremely high resolution observations of the KI profile
may help, and these were underway at the AAT in June 1994;
very preliminary results will be discussed in the epilogue. Although this
explanation is attractive, it would however not explain the atypical LiI/KI
ratio
observed in cloud B: it was shown that the LiI/KI ratio is rather constant
around 3.10$^{-3}$ in the ISM (White, 1986), as it is observed in cloud A for
instance, whereas that in cloud B is $\simeq0.03$. There is no reason
why a third absorbing cloud would explain this abnormal ratio since the LiI
column density in cloud B is derived mainly from the \li$_B$ doublet, which
could not be contaminated by the \li$_C$ line. Finally, although our
combination S/N--\res~did not allow to look with confidence for more than two
absorbing clouds, we nevertheless tried 3 cloud solutions in different
configurations and in each case, there was no way to escape having one of the
\iso ratios around 2--3.

The only standard way to obtain a ratio as low as $\sim$2 in the ISM comes
through a massive interaction of the GCR with the material of cloud B. Using
the
calculations of Reeves (1993), Steigman (1993), one may evaluate that
a burst of GCR spallation over 10$^6$\ts yrs is required with an
enhancement of the GCR flux by a factor $\zeta\sim2\times10^4$ to reproduce a
ratio of 2, and $\zeta\sim2\times10^5$ for a ratio 1.5. Such an
enhancement is so enormous that it would make \zo~a $\gamma$--ray source with a
flux detectable above the present instrumental thresholds, which has not been
reported as yet. Still, this would explain in a natural way the LiI/KI ratio
measured in cloud B, as in this scenario one should expect to measure a ratio
(LiI/KI)$_B\sim(1+10^{-4}\zeta)$(LiI/KI)$_A$, the factor 10$^{-4}$ arising from
the ratio of the duration of the burst to the age of the Galaxy. This
scenario seems unrealistic as for now, but one should note that the
calculations
are extremely coarse in that they assume a magnification of the whole GCR
proton
flux spectrum, and not simply the presence of a low-energy excess for
instance (Meneguzzi et al., 1975).

\vbox{
\epsfysize= 8 cm $$\epsfbox[72 130 550 600]{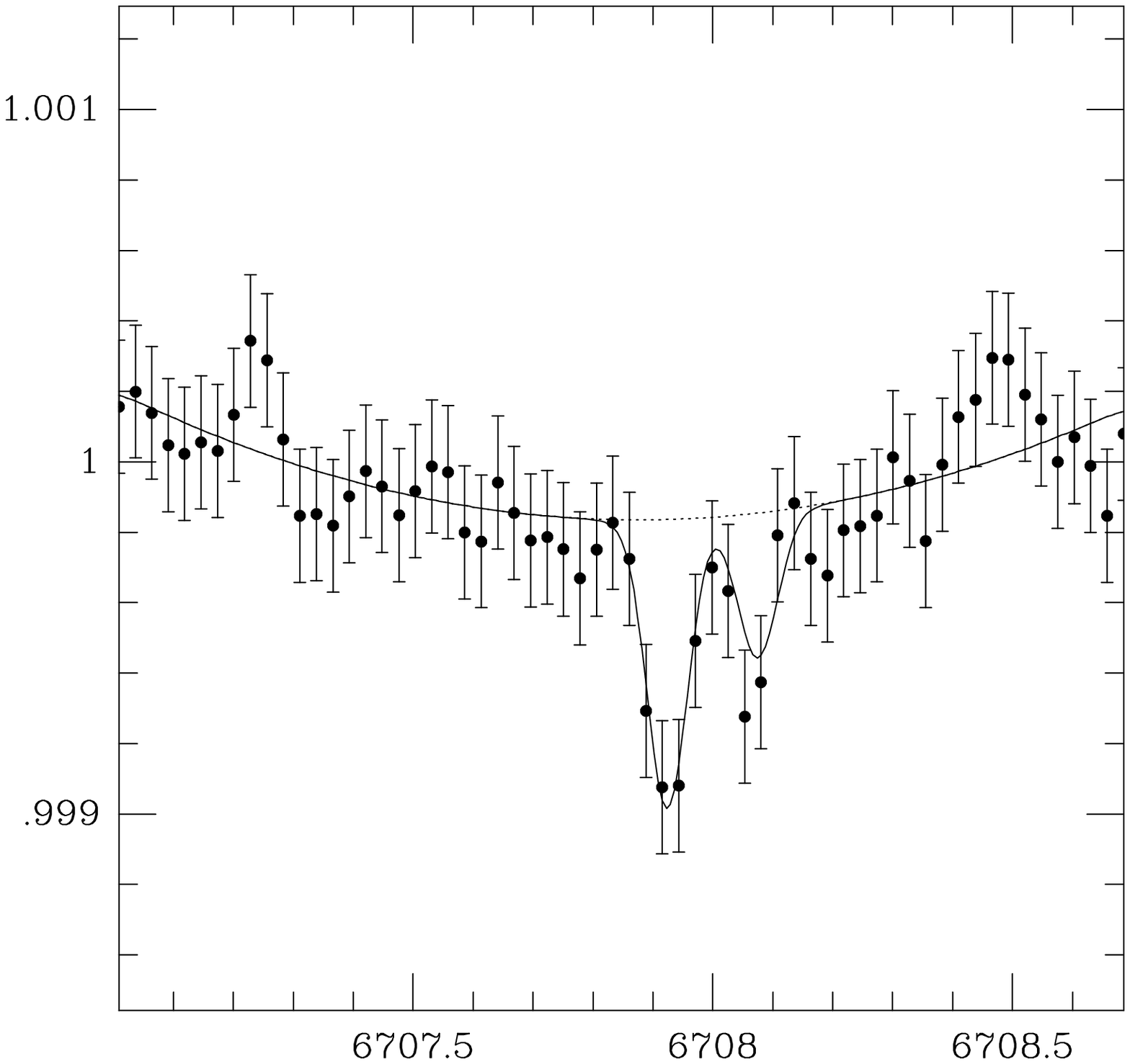}$$

\noindent{\ninerm {\ninebf Fig.4:} After subtracting all calculated
contributions
to the observed profile of Fig.3 except that of the \lis$_A$ doublet, this
latter
is shown here as residuals. The solid line shows the fit to this doublet
calculated in Fig.3.}
\bigskip}

\vbox{
\epsfysize= 8 cm $$\epsfbox[72 130 550 600]{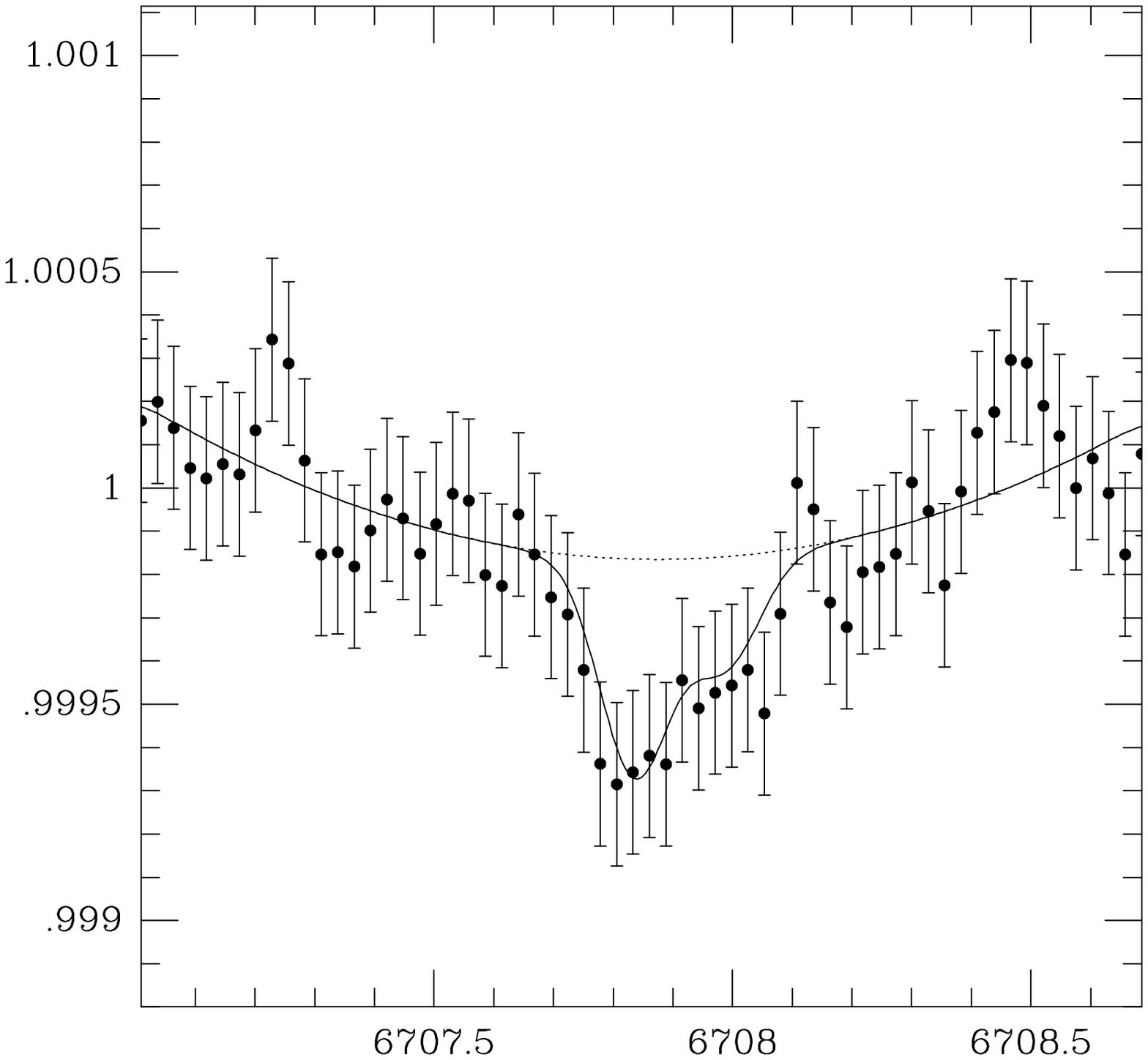}$$

\noindent{\ninerm {\ninebf Fig.5:} Same as in Fig.5, but for the \lis$_B$
doublet.
Note the scale in ordinates.}
\bigskip}

Of course, the \isop$_A$ ratio now depends on the suggestions presented to
account for the \isop$_B$ ratio, although none of them appears entirely
satisfactory. In each case however, one should expect that
the actual \isop$_A$ ratio be higher than what is measured, viz.
\isop$_A\ga8.6$.
Our final results toward \zo~are thus:
$$ \left(^7{\rm Li}/^6{\rm Li}\right)_A\,\ga\,8.6$$
$$ \left(^7{\rm Li}/^6{\rm Li}\right)_B\,\sim\,2\;({\rm unexplained?})$$
Very recently, Cass\'e,
Vangioni-Flam \& Lehoucq (1994) offered a very elegant explanation to the
\isop$_B$ ratio. They have interpreted the recent detection of an extremely
high
$\gamma$--ray flux in Orion as due to the interaction of accelerated $\alpha$
and
$^{16}$O nuclei with the ISM.
They showed that a SNII exploding inside an interstellar cloud would indeed
accelerate such nuclei, which are typical yields of a massive star, and
through interaction with
the ISM would produce \li~and \lis~isotopes, among others, with a production
ratio
\iso$\simeq3$ depending on certain parameters. It happens that although this
interaction process should last
$\Delta{\rm t}\simeq10^5$ yrs, its efficiency in creating Li atoms is some
10$^6$
times that of usual GCR spallation, and this is precisely the
$\zeta\times\Delta{\rm t}$ factor we were looking for above. This efficiency
results from the very low thresholds of $\alpha$--$\alpha$ reactions, the
shape of the energy spectrum, and the
weak ionization losses of $\alpha$ nuclei when propagating in the ISM. Assuming
now that cloud B is a fragment of an interstellar cloud irradiated {\it via}
this process, one should expect to measure \isop$_B\sim3$, and
(LiI/KI)$_B\simeq11$, which is indeed very satisfying!

 Finally, Crane \&
Lambert (private communication) obtained lithium data at the MacDonald 2.7m in
the direction of \zo, with S/N$\simeq2000$ and \res=1.25$\times10^5$. They
report the presence of two interstellar components with \iso ratios above 10 in
each, though the secondary component is endowed with a thermal width
b$\sim13$\ts km.s$^{-1}$, i.e. far above what is typically observed (a few
km.s$^{-1}$). Our solution also provides a good fit to their data, which in
turn argues against an instrumental contamination.
\vskip 1.0 cm
\no{\bf 4. Conclusions \& Epilogue}
\m
Previous measurements of the interstellar \iso ratio were discussed and
criticized in turn, and new data on an already observed line of sight were
presented. At the end, the observational status of this important isotopic
ratio
restrains itself to the following:
$$ \left(^7{\rm Li}/^6{\rm Li}\right)\,=\,11.1\;\pm2\;\;(\rho\,{\rm Oph})$$
$$ \left(^7{\rm Li}/^6{\rm Li}\right)\,\ga8.6\;\;(\zeta\,{\rm Oph})$$
$$ \left(^7{\rm Li}/^6{\rm Li}\right)\,\sim2\;\;(\rho{\rm ,}\,\zeta\,{\rm Oph,
\,to\,be\,confirmed?})$$
The measurements of MHW in the directions of \zp~were shown to be biased
because
all the detected ISM components were not taken into account in the profile
fitting. The lower limit \iso$\ga25$ obtained by FD toward \zo~is unexplained.
A \iso ratio in the secondary cloud detected toward \ro~has to be
further discussed as it was done for the \iso=2 ratio toward \zo. This latter
is
not readily explained, but might be the result of a new spallation process
proposed by Cass\'e, Vangioni-Flam \& Lehoucq (1994), where $\alpha$ and
$^{16}O$ nuclei accelerated in SNII would interact with the ISM to create LiBeB
isotopes, among others. First calculations show that this process is much more
efficient in creating the LiBeB isotopes than usual GCR spallation and
compensates largely for the short duration time of the interaction. In any
case,
this process is to be elaborated on further and cannot be neglected any longer
in studies of the galactic evolution of the light elements.

New observations of typical \iso target stars were conducted in June 1994 at
the
Anglo-Australian Telescope using the newly commissioned Ultra High Resolution
Facilities spectrograph at \res=6$\times10^5$. Toward each of the observed
targets \ro, \zo, and others, more than two interstellar components seem to be
detected in KI, some of them separated by only $\simeq1$\ts km.s$^{-1}$... This
means that to derive an accurate description of the velocity structure of a
studied
line of sight, which is essential as shown in 2., it is in fact mandatory to go
to a resolution \res$\ga 3.10^5$. What this entails for the \iso ratios is far
more depressing: there is no hope in LiI to resolve ISM components separated by
$\Delta {\rm v}\simeq1$\ts km.s$^{-1}$ because LiI is precisely a light
element,
and the natural width of the LiI line is thus already $\simeq1.2$\ts
km.s$^{-1}$.
Whenever the \iso ratio is around 2 in one component and 20 in the other, the
resulting profile will be indistinguishable from a profile obtained from a
single
absorbing cloud with a ratio \iso$\simeq10$... This means as well that the
above
\iso ratios are average values of \iso ratios in clouds with very
similar radial velocities $\Delta{\rm v}\simeq1$\ts
km.s$^{-1}<\,$\res.
One may however give the following considerations.
First of all, obtaining a representative value of the interstellar \iso ratio
is
no longer a matter of a few measurements. Just as for the D/H ratio on QSO
lines
of sight, it is a matter of statistics at a long term, and it certainly
deserves
the
effort. If variations of the \iso ratio are detected, then they must be greater
in
reality than what is observed, for the values measured are already averages
over
different absorbing clouds.  This would be the trace of an inhomogenous
physical
process of \li~creation, possibly the one proposed by Cass\'e, Vangioni-Flam
\& Lehoucq (1994).
If no variations are detected, we may have got the right value of the
interstellar \iso ratio, the system is homogeneous, and after all, one may
consider that two interstellar clouds separated by only 1\ts km.s$^{-1}$ are
somehow physically linked and that they should exhibit similar \iso ratios. The
epilogue of this review is thus that we only see the beginning of it. Stating a
ratio \iso$\sim10$ as representative of the ISM is an attractive
possibility, but it is a bit too premature.
\m
\no{\it Acknowledgements: the work reported of above was done in
collaboration with Alfred Vidal--Madjar.}
\b
{\bf References}
\m
\no Abia, C., Isern, J., Canal, R.: 1993, A\&A {\bf 275}, 96

\no Abia, C., Isern, J., Canal, R.: 1994, A\&A to appear

\no Cass\'e, M., Vangioni--Flam, E., Lehoucq, R.: 1994, Nature, to appear

\no Ferlet, R., Dennefeld, M.: 1984, AA {\bf 138}, 303 (FD)

\no Lemoine,  M., Ferlet, R., Vidal-Madjar, A., Emerich, C., Bertin, P.: 1993,
A\&A {\bf 269}, 469 (L93)

\no Lemoine, M., Ferlet, R., Vidal-Madjar, A.: 1994, A\&A, to appear

\no Meneguzzi, M., Audouze, J., Reeves, H.: 1975, A\&A {\bf 40}, 99

\no Meyer, D. M., Hawkins, I., Wright, E. L.: 1993, ApJ {\bf 409}, L61 (MHW)

\no Mowlavi, N.: 1994, {\it in ESO/EIPC Workshop on ``Light Element
Abundances''}, Isola d'Elba, ed. P. Crane, Springer--Verlag, in press

\no Pinseonneault, M. H., Deliyannis, C. P., Demarque, P.: 1992, ApJ Supp. {\bf
78}, 179

\no Reeves, H.: 1993, A\&A {\bf 269}, 166

\no Reeves, H.: 1994, Rev. Mod. Phys. {\bf 66}, 193

\no Smith, M. S., Kawano, L. H., Malaney, R. A.: 1993, ApJ Supp {\bf 85}, 219

\no Spite, M., Molaro, P., Fran\c{c}ois, P., Spite, F.: 1993, AA {\bf 271}, L1

\no Thomas, D., Schramm, D. N., Olive, K. A., Matthews, G. J., Meyer, B. S.,
Fields, B.: 1994, ApJ {\bf 430}, 291

\no Vauclair, S.: 1988, {\it in ``Dark Matter''}, ed J. Audouze and J. Tran
Thanh Van, p. 269

\no Welty, D. E., Hobbs, L. M., Kulkarni, V. P.: 1994, ApJ, to appear

\no White, R.E.: 1986, ApJ {\bf 307}, 777
\end